\documentclass{aa}

\usepackage{amsmath}
\usepackage{graphicx}
\usepackage{color}
\usepackage{ulem}

\usepackage{txfonts}
\usepackage{epstopdf} 
\usepackage{siunitx}
\usepackage{caption}
\usepackage{booktabs}  

\usepackage[draft]{hyperref}

\usepackage{array}
\newcolumntype{L}[1]{>{\raggedright\let\newline\\\arraybackslash\hspace{0pt}}m{#1}}
\newcolumntype{C}[1]{>{\centering\let\newline\\\arraybackslash\hspace{0pt}}m{#1}}
\newcolumntype{R}[1]{>{\raggedleft\let\newline\\\arraybackslash\hspace{0pt}}m{#1}}

\newcommand{\ctiPixel}{\texttt{CtiPixel}\ }

\begin{document} 

\title{\textit{Gaia} serial CTI modelling and radiation damage study}

\titlerunning{\textit{Gaia} serial CTI modelling and radiation damage study}

\authorrunning{Pagani et al.}

\author{
C. Pagani \inst{\ref{inst:ul}}\fnmsep\thanks{Corresponding author: C. Pagani\newline
e-mail: \href{mailto:cp232@le.ac.uk}{\tt cp232@le.ac.uk}}
\and N. C. Hambly\inst{\ref{inst:uoe}}
\and M. Davidson\inst{\ref{inst:uoe}}
\and N. Rowell\inst{\ref{inst:uoe}}
\and C. Crowley\inst{\ref{inst:hespaceesac}}
\and R. Collins\inst{\ref{inst:uoe}}
\and F.~van Leeuwen\inst{\ref{inst:ioa}}
\and G. M. Seabroke\inst{\ref{inst:mssl}}
\and A. Holland\inst{\ref{inst:ou}}
\and M. A. Barstow\inst{\ref{inst:ul}}
\and D.~W.~Evans\inst{\ref{inst:ioa}}
}  

\institute{
School of Physics and Astronomy, University of Leicester, University
Road, Leicester LE1 7RH, United Kingdom
\label{inst:ul}
\and
Institute for Astronomy, School of Physics and Astronomy, University of Edinburgh, 
Royal Observatory, Blackford Hill, Edinburgh, EH9~3HJ, 
United Kingdom
\label{inst:uoe} 
\and
Institute of Astronomy, University of Cambridge, Madingley Road, Cambridge CB3~0HA, United Kingdom
\label{inst:ioa}
\and
HE Space Operations BV for ESA/ESAC, Camino Bajo del Castillo s/n, 28691 Villanueva de la Ca{\~n}ada, Spain
\label{inst:hespaceesac} 
\and
Mullard Space Science Laboratory, University College London, Holmbury St Mary, Dorking, Surrey, RH5 6NT, United Kingdom
\label{inst:mssl} 
\and
The Open University, School of Physical Sciences, Centre for Electronic Imaging, Milton Keynes, United Kingdom
\label{inst:ou}
}

\date{}

\abstract{

   {}
   {}
   {During the course of its mission, ESA's \textit{Gaia} spacecraft has generated a map of the stars of the Galaxy of exquisite detail.
   While in its L2 orbit, the satellite has been exposed to high energy cosmic rays and 
   solar particles, that caused permanent damage to its CCDs.
   The main effect of radiation damage on \textit{Gaia} data is the distortion of its images and spectra, caused by 
   the CCDs charge transfer inefficiency (CTI) during the readout process, that, if not taken into account, 
   can result in inaccurate measurements of a star's location and flux.
   In this work, the impact of CTI in the serial readout direction, larger than in the parallel due 
   to the presence of CCDs manufacturing defects, has been analysed and modelled.
   A pixel-based, physically motivated CTI model, \ctiPixel, has been developed to characterise
   the damage in \textit{Gaia} CCDs. The model has been calibrated using dedicated serial CTI diagnostic data, 
   taken every 3-4 months over the course of the mission.
   The model is shown to be a good representation of the observed signatures of CTI in the calibration datasets, 
   and its parameters reveal significant insights into the nature of the CCD defects generated by space irradiation. 
   The evolution of the damage in the serial direction shows a general small linear increase over time, 
   with sudden step changes after strong solar flares and coronal mass ejections directed towards Earth.
   The serial CTI showed a further step increase as a consequence of the engineering CCD annealing experiment 
   carried out after the completion of \textit{Gaia} science observations.
    }

   {}  
  {}
}

   \keywords{instrumentation: detectors -- methods: data analysis -- space
vehicles: instruments}

   \maketitle
   \nolinenumbers

\section{Introduction}
\label{sec:intro}
The \textit{Gaia} satellite \citep{2016A&A...595A...1G}, ESA's astrometric mission, was
launched in December 2013 and collected data in its main science mode until January 2025,  
with the aim of creating the most detailed map of the stars of our Galaxy.
The spacecraft operated in the vicinity of the second Lagrange point (L2), 1.5 million km from Earth, scanning the sky 
combining observations from the field of view of two  telescopes, separated by a basic angle of 106 degrees, to measure the stars' positions, distances and 
proper motions with an accuracy of micro-arcseconds. In addition to the astrometric measurements, \textit{Gaia} performed 
photometric observations of all
detected sources, low-resolution spectrophotometry with the Blue and Red Photometers 
(BP and RP, and more generally referred to as XP devices) 
as well as near-infrared medium resolution spectroscopy with the Radial-Velocity Spectrometer (RVS) 
for objects brighter than $G_\mathrm{RVS}\sim16$.

\textit{Gaia}'s focal plane consists of a grid of 106 scientific e2v Charge Coupled
Devices (CCDs) -- see \cite{2016A&A...595A...6C}. 
The detectors have been exposed to a high-energy particle environment since
launch.
The characteristics of this irradiation, 
based on measurements using \textit{Gaia} detectors in combination
with data from other spacecrafts in L2, is described and analysed in Crowley (in prep).
Over time, space irradiation causes permanent damage to the detectors' silicon structure.
The defects that are thus generated are also referred to as traps, as they can cause the capture of electrons from a source signal 
and their release at a later time during the CCD readout process, distorting the acquired images.
This effect is quantified in terms of Charge Transfer Inefficiency (CTI), defined as the fraction of charge
lost to traps during the transfer of the signal from a pixel to the next (see, for example \cite{Hopk96}).
This was identified early on in the \textit{Gaia} development phase as a critical aspect of the mission. For the fainter stars 
in particular, CTI can cause a distortion in the observed images that strongly affects the 
derivation of the stars detector position and flux.

After launch, the impact of CTI on \textit{Gaia} data was found to be lower than originally anticipated
by early radiation model studies, as shown in \cite{2016SPIE.9915E..0KC}.
This was due to a number of factors. For instance, the timing of the launch, past solar maximum, 
and the low activity of the past solar cycle, proved beneficial, in terms of total irradiation dose. 
Another mitigating effect on CTI for \textit{Gaia} has been 
the overall higher than expected diffuse optical background component, 
that reduces the impact of CTI during readout, as its signal can keep some traps permanently occupied. 
In particular, in the Along Scan (AL) direction, the accumulation of damage during the mission resulted in a parallel CTI 
that remained a factor of $\sim$~8 below predictions. 
The serial CTI (sCTI) in the Across Scan (AC) direction 
has slowly increased over the years and in absolute terms is higher than in the parallel direction. 
The major cause of sCTI is the manufacturing traps. These, already present at launch, 
affect the serial transfers to a larger degree than in the parallel direction because of the faster
serial readout timing.

The focus of this paper is the analysis of the serial CTI effects on a set of \textit{Gaia} engineering data,
that are taken periodically during the course of the mission, and the mapping of the damage evolution over time.
In Section~\ref{sec:gaia_focal_plane_env} we describe the telescope focal plane and its detectors, 
the L2 radiation environment and the type of damage
it causes on the CCDs. We will review the preparatory work carried out before the satellite launch by
the \textit{Gaia} irradiation study team, aimed at predicting and modelling the effects of radiation in \textit{Gaia} data.
That analysis has informed the development of the \ctiPixel CTI model and software package 
described in this paper.  
The model is based on physical properties of the defects 
that cause CTI and their effects on charge during the detectors readout process.
\ctiPixel has been calibrated using a collection of 
dedicated serial CTI engineering activities, carried out periodically during the course of the mission,
that involved the controlled injection of charge at the top of the CCDs. 
These engineering tasks and calibration datasets are presented in 
Section~\ref{sec:gaia_scti}, while the \ctiPixel model is described in Section~\ref{sec:gaia_cti_model}.
In Section~\ref{sec:trails_fits} the method used to process the serial calibration activities and to fit them with
the \ctiPixel model will be shown. The results of the fits are presented in Section~\ref{sec:results}.
Finally, the physical interpretation of the fitted trap parameters, the limitations of the model 
and an attempt to use \ctiPixel to treat the signatures of serial CTI 
in \textit{Gaia} science observations are discussed in Section~\ref{sec:discussion}. 
Appendix~\ref{appendix:eol} presents the main effects on serial CTI following the CCDs annealing procedure, 
one of \textit{Gaia} End-of-Life activities. 

\section{\textit{Gaia} CCDs and radiation environment}
\label{sec:gaia_focal_plane_env}

The 106 detectors that form the \textit{Gaia} focal plane are categorized by their
function and properties, with the majority of them (62) identified
as Astrometric Field (AF) CCDs, dedicated to the measurement of the position of the
stars detected by the spacecraft Star Mapper (SM) devices -- for a full description
of the \textit{Gaia} focal plane see \cite{2014SPIE.9154E..06K}.
All detectors have image section dimensions of 4500 by 1966 pixels in the
parallel and serial readout directions, with each pixel the physical size of $ 10 \times 30$~$\mu$m. The main
difference between the devices employed in the focal plane is their
thickness.
For the AF and Blue Photometers (BP) CCDs the thickness is 16~$\mu$m, while for the red variants employed in the
RP and RVS detectors a higher value of 40~$\mu$m enhances their sensitivity for sources emitting
mainly at longer wavelengths.

\subsection{CCD operations}
\label{subsec:ccdops}

The impact of radiation damage on the images and spectra acquired by a CCD
depends both on the characteristics of the device and on how 
it is operated, for example on the timings of the readout process
and the CCD temperature.
A critical aspect for \textit{Gaia} is the way its images are acquired, exposing the CCDs
continuously while they are clocked in the parallel direction.
This technique is known as `Time Delayed Integration' (or TDI) mode, in which
the detectors are clocked in the parallel direction with a period of 0.9828 millisecond (corresponding to a clocking frequency of 1.017~kHz), 
matching the scanning rate of the stars over the detectors.
Therefore, the signal from a transiting source is accumulated during
the readout, in contrast to the typical CCD acquisition technique of exposing
the camera at a fixed sky position for a determined amount of time 
followed by the readout of the frame.

The parallel transfer rate in TDI mode is rather
long compared to that in more common CCD imaging. Furthermore, this is in stark
contrast to the two-phase readout serial clocking frequency, which is extremely fast (10~MHz
corresponding to charge transfer times of~0.1~$\mu$s in the serial register) except when
samples are being read, yielding longer transfer times of order~10~$\mu$s upstream in the
serial register.
The fast--slow serial scan is further complicated by a readout suspension around
the four--phase parallel clocking. This is because the video chain electronic offsets
would be otherwise hugely disturbed during readout sampling by the voltage
swings commanding those parallel phases --- residual offset disturbances on resumption of the serial
readout are presented in \cite{DR2-DPACP-29} where these aspects are
discussed in greater detail. 
At each readout pause, the dwell time in the serial register, during which charge is confined 
under a pixel before being moved to the succeeding one is~80.4~$\mu$s 
for AF devices, and~76.8~$\mu$s for XP CCDs. Hence the
serial scan consists mainly of about 2000 transfers of~0.1~$\mu$s, within which 
are scattered several tens of dwell times of~10~$\mu$s in contiguous groups
(depending on instrument mode, allocation of science windows and of
sacrificial `braking' samples used to further sample away the offset disturbances)  
plus three dwells of around~80~$\mu$s (one readout pause having taken place
before the serial scan starts).

There are three other important characteristics of the \textit{Gaia}'s CCDs that
influence the impact of irradiation on its measurements.
\textit{Gaia}'s devices were designed and manufactured with pixels presenting a
supplementary buried channel \citep[SBC;][]{2013MNRAS.430.3155S}, that
limits the volume within a pixel where the charge signal from a source is
enclosed during readout, thereby reducing parallel CTI effects (the serial register has no
SBC).
The second is the charge injection functionality of the CCDs, in which a
controlled amount of charge is injected at the top of the devices at specific
intervals. During observations taken in normal science mode, 
a block of four contiguous lines are injected with approximately ten
thousand electrons of charge per pixel every~2000 (AF mode) or~5000 (XP mode) lines.
When these blocks of injected lines are  readout they temporarily fill the
defects of the entire CCDs with a fraction  of their charge, mitigating the
effects of CTI.
Finally, of particular importance for CTI effects is the detectors' temperature. 
On \textit{Gaia}, the CCDs nominal operating temperature is~163~K, 
with some variations of several degrees observed over the focal plane. 
The temperature has been found to be stable over time, albeit
with low--level seasonal variations dependent on heliocentric distance 
in addition to abrupt variations in short intervals during specific 
mission phases such as the decontamination events and other disturbances to 
power dissipation in the focal plane array caused by operational changes (see
Section~\ref{subsec:finalcalibs}).

\subsection{Radiation environment at L2}

The radiation environment experienced by \textit{Gaia} at L2 is the focus of Crowley (in prep).
Its analysis is based on the data acquired by the 14 Star
Mappers, the devices dedicated to the stars' on-board detections. In particular, as part of the automated detection algorithm, 
Prompt Particle Events (PPEs) are identified. 
These are features in the SM images that are too sharp in comparison 
to the \textit{Gaia} Point Spread Function shape to be considered true signals from point-like sources, and are instead attributed to 
high-energy particles, mostly protons, either cosmic rays or originating from the Sun. The observed PPEs rate can therefore be considered
as a real-time measure of the radiation environment experienced by \textit{Gaia}. 

The impact of high energy particles on the detectors can cause 
ionizing or non-ionizing  radiation.
In the first instance irradiation is revealed over time as shifts in the measured CCDs baseline voltage values \citep{90114}, that can be 
compensated by hardware setting adjustments during the course of the mission.
The non-ionizing radiation instead consists of the impact of high energy particles with the silicon crystal of the detectors, 
that can lead to the displacement of one or multiple atoms from their original positions in the lattice structure, and
is therefore also known as displacement damage \citep{4323299}; 
the atoms knocked out of their original sites 
leave a vacancy, that over time can be filled by impurities already present in the detector, such as Phosphorous or Oxygen atoms \citep{PhysRev.134.A1359}, or
form a stable bonding with another vacancy (a type of defect called a Divacancy, \cite{45376}). 
These defects can become permanent, generating spurious energy levels between the valence and the conduction band of the semiconductor, 
where electrons can become temporarily trapped
during the CCD readout process. The amount of time after which a trapped electron is released back into
the conduction band is called the release (or emission) time, and it characterises the defect species, as it is
a function of the specific trap energy level and the temperature at which the camera is operated. 
The trapping of electrons during the signal readout in CCDs is the cause of CTI, 
quantified as the fractional loss of charge during the transfer between two pixels.
Common defects of silicon devices with their characteristic energy below the conduction band are listed in Table~\ref{table:defects}.  

\begin{table}
\caption{Common silicon defect species}              
\label{table:defects}      
\centering                                     
\begin{tabular}{c c c c}          
\hline\hline                        
Type & Symbol & Energy (eV)  \\    
\hline                                   
    Phosphorous & Si-E & 0.44  \\      
    Divacancy & V-V & 0.41       \\
    Unknown &  & 0.34         \\
    Multi-vacancy & V-V-V & 0.20       \\
    Oxygen & Si-A & 0.17       \\
\hline                                             
\end{tabular}
\end{table}

\begin{figure}
  \centerline{\includegraphics[scale=0.45]{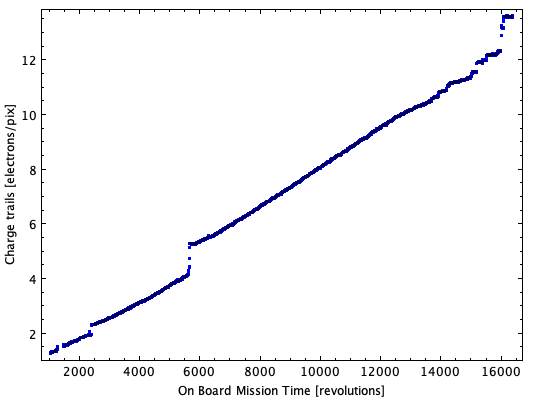}}
   \caption{Evolution of the measured charge trails intensity in the parallel direction averaged over all Astrometric Field devices,
    covering the entire course of the \textit{Gaia} mission.
   The trails are an empirical measure of the amount of radiation accumulated in orbit. The gradual increase in damage is caused by 
   Cosmic Rays, while strong solar flares directed towards Earth are responsible for the step changes in the damage evolution.
   Time intervals affected by temperature disturbances of the focal plan, due for example to decontamination activities, 
   or instances of incomplete telemetry have been removed from the plot for clarity.}
              \label{fig:ptrail}
\end{figure}

On board the \textit{Gaia} spacecraft, the parallel CTI level and its evolution are monitored 
using the same blocks of Charge Injection (CI) lines 
that are used to mitigate the effect of radiation damage.
The pixels following the blocks of CI lines present characteristic
charge trails, that are caused by the release of previously captured 
electrons in a stochastic process.
The total charge in the trails provides a measurement of the number of active defects in the detector, 
and is a proxy for the CTI in the parallel direction. Figure~\ref{fig:ptrail} shows the measured intensity 
of the parallel trails, averaged over all the AF CCDs, for the entire span of the \textit{Gaia} mission, with time 
expressed in units of On Board Mission Time (OBMT) revolutions, corresponding to approximately one satellite rotation \citep{DR2-DPACP-51}.
In the serial direction, a similar analyses is performed using 
the charge trails in the serial register post-scan pixels
measured during dedicated serial CTI calibration data sets.
These special engineering activities are described in detail in Section~\ref{sec:gaia_scti}.

The analysis of the charge trails in \textit{Gaia} CCDs has revealed a number of significant results
that were first reported by \cite{2016A&A...595A...6C} using data
from the initial two years of the mission. 
The measured evolution of the damage is characterised by a steady, gradual increase 
since the beginning of the mission, that is attributed to the
continuous irradiation from the Galactic Cosmic Rays (CR) flux (mostly protons and alpha particles). 
A second component of the damage evolution is seen as sudden steps at specific
times in correspondence with significant solar activity events. 
The strongest of these occurred on the 10th of September 2017 (this correspond to an 
On Board Mission Time of 5651 revolutions), 
that saw a strong class X8.2 solar flare emitted in the Earth's direction
\citep{Seaton_2018}. 
The high energy particles from that flare caused the generation of new defects in the detectors
that resulted in a sudden increase of approximately 20\% in the level of the charge trails.

A wealth of additional information can be derived from the analysis of the trails. 
For example, it can be seen that the gradual increase in the damage 
follows the 11-year modulation of the solar cycle, with a steeper accumulation rate during solar minimum and a shallower increase
at solar maximum, in correspondence with the higher and lower levels of the measured Galactic CR flux. 
Another feature of the observed damage is its non-uniformity
over the focal plane, due to the different amount of shielding from the irradiation and the mitigation effect 
of the local diffuse background signal.
The analysis of the charge trails can reveal detailed information on the physical nature of the traps,
such as the characteristic emission timescale of a specific defect type. 
This will be described and discussed in the next 
Sections~\ref{sec:gaia_scti} and~\ref{sec:gaia_cti_model}.

\section{\textit{Gaia} serial CTI calibration activities}
\label{sec:gaia_scti}

The impact of radiation on the CCDs serial readout efficiency has been monitored throughout the course
of the \textit{Gaia} mission with a set of recurrent, dedicated  serial CTI calibration tests.
These engineering activities, repeated every three to four months, comprise the
injection of blocks of consecutive lines with charge at several intensity
levels, regulated by the CCD injection voltage setting. 
At the lowest setting, the blocks are filled with approximately
$10^3$~electrons/pixel.
During the activity, the voltage is raised a total of five times,  
reaching an injection level of up to a few hundred thousands of electrons per pixel.
The full procedure results in five sets of data of increasing charge injection
flux, each one of the five sets consisting of 70 blocks of 225 consecutive
charge injected lines, followed by 200 lines with no injected charge.
In a line, the injected charge value is not exactly the same in all pixels, 
but the charge pattern repeats line by line. 
Moreover, within a block, in spite of a constant commanded injection level, 
the actual signal falls slightly (a phenomenon known colloquially as `injection
droop').

When these engineering data are read out in the serial direction and the output node,  
the flux level of the final 12 pixels of the injected lines is recorded,
along with the signal of 20 post-scan samples (24 for AF1 CCDs, 
the first strip of seven detectors of the AF devices).
The post-scan pixels are `virtual' pixels, read out by over-clocking the serial
register beyond the last image section column.
Their signal follows an exponential decay (a trail), formed by 
deferred (or released) charge from previously 
trapped electrons, superposed on the electronic bias level of the video chain.
The cumulative amount of signal in the post-scan samples, after subtraction of the 
CCD electronic offsets (the electronic bias),
is an indicator of the density of traps in the serial register, 
and therefore of the amount of total irradiation and damage of the devices that
is affecting the serial transfer process. 

\begin{figure*}
  \centerline{\includegraphics[scale=0.65]{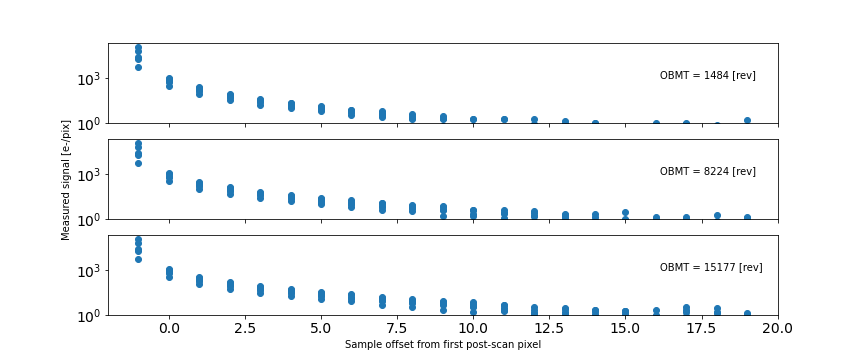}}
   \caption{Example charge trails from three epochs of \textit{Gaia} in--orbit serial
   CTI calibration activities. The mean level of the charge injection in the
   image section CCD columns is shown before the start of the post-scan
   measurements illustrating the 2 orders--of--magnitude range in the
   commanded blocks. A small increase in serial CTI is observed from early
   to late epochs in the mission time line, particularly for the release
   signature at longer timescales. These data come from device AF5 on row~5 near
   the middle of the focal plane array.}
              \label{fig:trailseg}
\end{figure*}

A total of~26 serial calibration activities has been carried out since the launch of \textit{Gaia}.
In Figure~\ref{fig:trailseg} we show as an example the measured deferred charge
levels in the post-scan pixels from three epochs of serial calibration activities, at 
each of the five charge signal levels, to illustrate the data input into the
subsequent analyses. 
These measurements reveal a number of important aspects in regard to the damage. 
The shape of the charge trails in the post-scan pixels is reminiscent of a sum
of exponential decays, regulated by the release timescales with which trapped charge is emitted from defect sites. 
Only specific trap species with an emission time of the same order as, or longer
than, the serial clock period contribute to the observed trails. 
Faster traps will release the charge just captured within the same pixel, while the
slower traps will release their charge on longer timescales compared to the CCDs serial readout frequency.
These data also display the effect of signal size on the amount of charge losses during readout. 
Larger amounts of injected charge occupy a bigger physical volume in the pixels, 
and can be affected by a higher number of traps, 
and lose more electrons during the charge transfers.
Finally, the charge trails grow larger at later times during the mission. 
The integral under their curves corresponds to the total amount of charge released 
by traps after a line is readout into the first~20 following pixels, 
and the increase over time is a consequence of the accumulation of damage.

A dedicated serial CTI model has been developed to process the
serial calibration activities. The model is physically motivated,
with the goal of extracting as much information as possible on the nature of the damage caused by space irradiation at L2.
The model, named \texttt{CtiPixel}, and its calibration based on the engineering test data 
are presented in the following Section.

\section{The \textit{Gaia} serial readout and CTI model, \ctiPixel}
\label{sec:gaia_cti_model}

In the years prior to launch, the \textit{Gaia} Data Processing Consortium set up the 
\textit{Gaia} Radiation Task Force, with the goal of developing 
the most effective model for the treatment of the effects of CTI on \textit{Gaia} images.
In the Task Force initial investigation, it soon became clear that to deal with the constraints posed by 
the very large volume of \textit{Gaia} observations and the available computational resources, 
the only effective way to handle CTI without introducing a noise component of unknown characteristics in the data 
was through a `forward modelling' approach, 
in which the model of a source signal is distorted using the adopted
CTI algorithm, and the derived damaged stellar profile is then compared to the measured flux
of the source. The first CTI modelling implementation proposed by the Radiation Task Force was completely empirical,
and was intended to be simple and therefore computationally fast, 
to be built upon over time with more detailed and realistic models, 
closer to the true physical nature of the damage.

The most sophisticated model that followed the initial approach 
was named `Charge Distortion Model 03', {\it CDM-03}.
In {\it CDM-03}, the proposed analytical model \citep{2013MNRAS.430.3078S} is
based on the physical theory of charge traps known as Shockley-Read-Hall (SRH, \cite{1952PhRv...87..835S}).
Two aspects of this model are particularly important. 
Firstly, the process of capture of electrons by the traps is not assumed to be instantaneous, 
as in the case of more traditional CTI models, but instead follows a probabilistic 
mechanisms, that is a function of the number of electrons in the signal. 
Moreover, the concept of a `density driven' regime to 
model the physical volume occupied by a signal charge cloud within a pixel
was introduced in {\it CDM-03}. This was in contrast with the classical 
`volume driven' CTI modelling, in which the occupied volume grows with the number
of electrons in the signal. 
In the density driven regime the signal occupies all the available pixel volume, 
no matter its size. This choice was made to account for the results observed in \textit{Gaia} irradiated devices before launch, 
that presented an unexpectedly large reduction 
in the measured CTI in the presence of a low, constant diffuse optical background. 
This was suggestive of the fact that the presence of a handful of electrons
in each pixel of the detector had the effect of keeping the defects filled, 
mitigating the effects of CTI. 

The CTI model developed for this work, named \texttt{CtiPixel},
inherits the fundamental principles on which the initial \textit{Gaia} Radiation Task Force approach 
and the {\it CDM-03} model
were based upon, and integrates them with tasks to closely reproduce the readout of charge signal
in the serial register of \textit{Gaia} CCDs.
In particular, it implements a `forward modelling' treatment of the CTI effects
and a physical characterisation of radiation damage, including
the stochastic nature of charge trapping and release.
It also  includes the modelling of the impact on CTI of the `scene', intended as the set of components present
in the image in addition to the source signal itself, such as the one from other nearby stars or from straylight. 
The additional components have an impact on dynamics of the traps statistics, 
as for example a diffuse background signal can fill traps and lower
the measured effects of CTI (this is known as the `sacrificial charge' effect). 
Finally, \ctiPixel adopts a pixel based implementation, modelling the charge losses, the transfer of charge
and the evolution of the trap statistics at the pixel level.

The choice of a physical model over a set of empirically derived functions has a number of important advantages. 
For instance, linking the model parameters to the physical properties of the damage  
makes their interpretation straightforward, giving insights into the nature of
the defects that are generated by space irradiation; moreover, the parameters derived from a physical model  
are independent of the CCD operating modes and settings, and can be directly applied to different data types and scenarios.
The choice of a pixel level implementation comes at the cost of a high demand
in computational resources and processing time,
but avoids simplifications and  approximations, providing a benchmark CDM version
of the highest accuracy.

\ctiPixel has been employed to process the datasets acquired during the serial calibration activities 
that were described in Section~\ref{sec:gaia_scti}.
When the signal in the serial register line is processed the readout is simulated assuming an initial  
uniform charge injection flux over the entire serial register,  calculated
using a robust estimate of the measurements of the CI lines in a block, 
corrected by the residual bias evaluated from the pre-scan pixels.
The sequence of serial transfer times for a specific CI line is retrieved using an utility that provides 
the dwell times (see previously) for the full readout process.

The modelling of the readout starts by shifting the charge by one pixel toward the readout node.
For every pixel of the serial register, the \ctiPixel processing task
evaluates the charge losses due to the traps and the number of electrons released 
by each trap species into the pixel ($L_i$, and $R_i$, respectively, for the $i_{th}$ trap species), 
and updates the total number of electrons $F$ 
in the signal at pixel $x$ accordingly:

\begin{equation}
F = F(x) + \sum_{i} (R_i - L_i)
\label{equ:flux_update}
\end{equation}

In \ctiPixel the capture and the release of electrons by traps is modelled as stochastic processes,  with a 
capture probability $P_c$ and emission probability $P_e$ specific to each trap type. 
The probability of a trap capturing an electron in a time interval $t$ 
is a function of the signal flux $F$ in the pixel:

\begin{equation}
P_c = 1- \mathrm{e}^{-\frac{t}{\tau_c}} = 1 - \mathrm{e}^{-\alpha t F^{\beta}}
\label{equ:Pcapture}
\end{equation}

The capture parameter $\alpha$ in the above equation includes the trap cross section $\sigma$, 
the thermal velocity of the electrons $v_{th}$, and the effective confinement volume $V_c$ of the signal. 
The capture time constant $\tau_c$ is such that commonly, in classical radiation damage algorithms,
the capture of an electron by a trap is assumed instantaneous ($P_c = 1$). 
In the case of \textit{Gaia}, it is instead necessary to compute this probability explicitly,
as its value can be smaller than one, in particular for the small charge packets of the faintest sources.

A defect can capture an electron if its location 
is within the physical volume occupied by the signal (its confinement volume $V_c$). 
The $\beta$ parameter in Equation~\ref{equ:Pcapture} models how $V_c$
varies as a function of the source signals size $F$, so that $\frac{V_c}{V_P} = (\frac{F}{FW})^\beta$,
where $V_P$ is the physical volume of the pixel and $FW$ its full well capacity, in number of electrons. 
A $\beta$ value of 1 is equivalent to the classic linear scaling of volume with number of electrons 
(the so called volume-driven model), while a $\beta = 0$ represents the regime 
in which a cloud of any size physically fills the entire pixel volume 
(the density-driven model mentioned earlier in the Section).

The capture probability of the trap species $i$, $P_{c, i}$, (Equation~\ref{equ:Pcapture}),
is used in \ctiPixel to calculate the charge losses in a pixel in conjunction with 
the following quantities:
the signal flux in the pixel $F(x)$, 
the fraction of the pixel's physical volume that the signal occupies 
($V_f = V_c/V_P$),
the density $\rho_i$ of each species,
and the fraction of empty traps ($f_{empty, i}$).
To summarize, the total charge losses $L$ in a pixel are therefore calculated as:

\begin{equation}
L = \sum_{i} L_i = \sum_{i} F(x) V_f \rho_i f_{\mathrm{empty},i} P_{c, i}
\label{equ:Ncapture}
\end{equation}

The probability of a defect emitting a previously trapped electron has a characteristics emission time ${\tau_e}$: 

\begin{equation}
P_e = 1- \mathrm{e}^{-\frac{t}{\tau_e}}
\label{equ:Pemission}
\end{equation}

that is a function of the specific trap energy level $E_t$, the defect cross section $\sigma_n$ and the CCD operating temperature $T$, expressed as:

\begin{equation}
\tau_e = \frac{\mathrm{e}^{\frac{E_t}{kT}}}{\chi \sigma_n v_{th} N_c}
\label{equ:tau_emission}
\end{equation}

where $\chi$ is the entropy factor ($\simeq 1$),  
$v_{th}$ the thermal energy of the electrons 
and $N_c$ are the number of effective states in the conduction band. 

The emission (or release) time varies greatly among trap types, as
its value is very sensitive to the characteristic energy level $E_t$ of the species.
Its measurement is therefore a very powerful method to identify the physical nature of the defect. 
Trap species like the 
O-Vacancy complex (or A-trap), with a characteristic energy of 0.17~eV or the
multi-vacancy (V-V-V trap) at $\simeq 0.20$~eV, 
are expected to have the greatest impact on serial CTI in \textit{Gaia} devices.
At the nominal \textit{Gaia} CCDs temperature of -110~C, using Equation~\ref{equ:tau_emission},
the emission times of these species are of the order of a few to hundreds of microseconds, 
close to the serial readout sampling value.
Other common species, like the Phosphorous-vacancy defect mentioned earlier, 
characterized by an energy level below the conduction band of approximately 0.44~eV, have 
emission times too long to affect the serial readout process. Once these defects capture an 
electron they remain filled for timescales much longer than the serial transfer process, and
are said to be effectively `frozen'. These species have no net impact on serial CTI.

In the first attempts at fitting the serial calibration data sets using \texttt{CtiPixel},
a fixed, nominal CCD temperature of -110~C was used for all the devices of the \textit{Gaia} field of view.
This was based on preliminary analysis of the spacecraft telemetry indicating a generally stable
temperature over the field of view. A more detailed investigation of thermal variations has
later shown differences of several degrees between devices, as well as seasonal and long-scale 
fluctuations. 
Variation of up to 5 degrees in temperature over the field of view can lead to 
uncertainties in the estimates of the release timescale of fast trap species as the A-trap of up to 30\% (see Equation~\ref{equ:tau_emission}).
The updated CCD temperature information has subsequently been introduced to the fitting model.
This change allowed the direct derivation of the characteristic energy level $E_t$ of the species
from the fits, in place of the temperature-dependent emission times ${\tau_e}$.

In \texttt{CtiPixel}, the amount of previously captured charge by traps that is released back into a pixel is determined for each
trap species by multiplying the release probabilities of Equation~\ref{equ:Pemission} by the number of traps 
filled by captured electrons. The total number of electrons released into a pixel is therefore calculated as:

\begin{equation}
R = \sum_{i} R_i = \sum_{i} \rho_i f_{\mathrm{filled}, i} P_{e, i} 
\label{equ:released_number}
\end{equation}

where $\rho_i$ is the number density of the $i_{th}$ trap species, $f_{filled, i}$ is the fraction of traps 
filled with a captured electron for that species, and $P_{e, i}$ its emission
probability.

\ctiPixel simulates the serial readout process by updating the signal flux $F(x)$ of every pixel
for the charge losses (Equation~\ref{equ:Ncapture}) and the charge released by 
each trap species (Equation~\ref{equ:flux_update}) during a serial readout pixel clocking cycle, 
and tracking the defects statistics
updating the number of empty and filled traps.
The signal is then shifted by one pixel towards the readout node, and the
calculations illustrated above are repeated, until the full readout register has been readout, 
and additionally evaluating the charge released by traps in the 20 (or 24 for AF1) post-scan pixels.
At this stage, the flux of the bottom line from the CCD image section is 
transferred into the serial register, and the process is repeated until the serial 
readout of the full calibration dataset has been completed.

The parameters of \ctiPixel that are fitted using the serial calibration activities data sets 
are the traps number densities $\rho_i$, 
the capture parameters $\alpha$ of Equation~\ref{equ:Pcapture} and the
volume scaling $\beta$ values, for each trap species. The parameter $\beta$ is allowed to change by species to account for 
possible variations in the spacial distribution of the different impurities in the silicon.
In addition, as mentioned earlier, during the initial fitting attempts the traps emission timescales $\tau_e$ 
(Equation~\ref{equ:tau_emission}) were also derived, employing the nominal \textit{Gaia} temperature value of -110~C 
for all devices and epochs. This choice was necessary as the complete telemetry from the temperature sensors over
the \textit{Gaia} focal plane wasn't available at the time the project was started and the initial model was being developed.
In the refined \ctiPixel implementation an epoch-dependent CCD temperature was 
used for each camera, derived from the spacecraft telemetry, allowing the direct fit of the
characteristic energy level $E_t$ of different species.

In the model, the trap species parameters $\tau_e$ (or $E_t$), $\alpha$ and $\beta$ are 
set to be independent of time, as they are considered intrinsic characteristics of a defect type, while the traps
number densities $\rho_i$ are allowed to vary to track the increase of defects in the CCD caused by radiation.
The capture parameter $\alpha$ can be converted into a characteristic capture cross section $\sigma$ of units $cm^2$ using the
physical properties of the \textit{Gaia} CCD pixels: their geometric volume $V_g$, 
the serial full well capacity $FW$ of 475K electrons, 
and the thermal velocity of the electrons ($v_{th} \approx 1.21x10^7cm/s$):

\begin{equation}
\sigma = \frac{2 \alpha  V_g} {v_{\mathrm{th}} \mathrm{FW}^{0.5}}
\label{equ:xsection}
\end{equation}

The surface area of a readout register pixel is equal to 30~$\mu m$ in the horizontal (H) direction, to match the width 
of the pixels in the image area, and 15~$\mu m$ in the vertical (V) direction, larger compared to CCD image pixels
\citep[for a detailed description of the \textit{Gaia} CCDs
architecture see][]{2005SPIE.5902...31S}.

This design allows serial pixels to handle a minimum of 2.5 times the full-well capacity of an image pixel. 
The depth (d) of the readout pixels was estimated 
at approximately 1~$\mu m$, based on pre-launch modelling and measurements by DPAC (private communication). 
The readout in the serial direction uses a two-phase scheme, 
therefore the signal packet cloud occupies one half of the pixel
width at any one time, resulting in a total geometric volume 
$V_g = \mathrm{H} \times \mathrm{V} \times \mathrm{d} = \frac{30}{2} \times 15 \times 1 \, \mu m^3 = 2.25x10^{-10} cm^3$,
In the initial attempts at fitting the serial calibration data sets
using \texttt{CtiPixel}, the number of trap species in the model was set to two, based
on prior knowledge that both the Oxygen and the multi-vacancy defects could affect the serial readout in \textit{Gaia} devices.
The first fits results showed that this choice was not sufficient; on the contrary, 
it was found that a nonphysical high number of trap species was required to achieve a satisfactory fit to the images.
This was interpreted as a shortcoming of the model, 
and was tackled by introducing an additional process to the readout sequence: 
the possibility of re-capturing the electrons just released by previously filled traps within the same pixel. 
Charge re-capturing can be activated by setting a flag in \ctiPixel.
When the flag is set, the released charge is added to the signal flux of a pixel before the capture probability is computed. 
Charge re-capturing showed to improve the best fit results, as will be presented in detail in Section~\ref{sec:trails_fits}.

The code in \ctiPixel has been optimised and a set of numerical approximations
was introduced to reduce the processing time of its computationally intense pixel-based implementation.
The most significant optimisation is relative to the treatment of the 
pixels with a common input flux and the same traps state of the neighbouring pixels. These cases are recognised and flagged
in the code so that the results of the model computations can be reused, avoiding unnecessary repetitions.

\section{Serial post-scan trails fitting}
\label{sec:trails_fits}

The serial calibration activities have been performed a total of~26 times 
during the mission, with the first run at On-Board Mission Time 
1484 revolutions, and a final 
acquisition at OBMT revolution~16381, on the 14th of January 2025, shortly 
before the end of \textit{Gaia} science operations. 
The data recorded from these activities consist 
of the measurement of the signal from a limited number of pixels:
the final~12 CCD image area pixels of each line, to
estimate the level of injected charge in the blocks, and the 
measurements in~20 (24 for devices in the first AF strip) post-scan pixels samples,
that show the characteristic charge trails from the release of trapped electrons in
the serial register. Data from all epochs have been fitted using 
the \ctiPixel serial readout modelling task.
 
Within a block, the injected charge level is only
approximately uniform, with the signal falling slightly from one line to the
next despite a constant commanded injection voltage, due to the
`injection droop' phenomenon. In our analyses, the signal from the~10$^{th}$
line in a block is used to estimate the amount of injected charge,
calculating the median of its values in the~70 blocks acquired at the same
injection setting. The signal in the post-scan pixels must be evaluated
precisely to detect levels of trailing charge
as low as 0.1 electrons per pixel, on average. 
To this end, particular care has been given
to accurately calibrate the electronic offsets, applying the
required bias non--uniformity corrections~\citep{DR2-DPACP-29}. The bias offset
level for each pixel in the serial readout sequence is characterised using line
scans made between the injection blocks with the first TDI blocking gate
\citep{2016A&A...595A...6C} in operation. Use of the gate ensures 
negligible photoelectric signal contamination to the offset values, but may
introduce gate--dependent effects relative to the trail measurements that are
made with no TDI gate active. The vector of offset values is measured relative
to the final post-scan sample and then applied in conjunction with the prescan
level on a line--by--line basis for the trail measurements. In this way any
global gate--mode dependent effect is removed, although there likely remains
some very low level of scan position dependent effects in the measurements. We
note that in AF mode the~12th sample (9th in XP mode) in the post-scan trail
sequences follows a readout pause and is subject to a large `glitch' offset \citep{DR2-DPACP-29}. Any
inadequacies in the electronic offset correction manifest as a discontinuity
in residual level from the~11th to the~12th trail sample in AF mode (8th to~9th
in XP mode); in general no such discontinuity is observed but may be visible for
one problematic device (see Section~\ref{sec:results}).

Two more lines of lower signal with respect
to those of the five charge injection levels commanded in the blocks 
were included in the \ctiPixel serial readout processing  and the fits,
with the goal of better constraining the model parameters.
The signal in the lines immediately after a block are dominated by the typical exponential charge trails 
from parallel CTI due to traps present in the CCD image area that release electrons during the parallel readout.
The first line of the release trail immediately following a block presents the
highest such deferred signal, and therefore its serial trail charge in the
serial post-scan pixels has the higher signal to noise when compared to the
other lines following a block, but this was not employed in the analysis to
avoid possible blooming charge contamination from the charge injection block.
Instead, the signal from the second line
following a charge injection block was preferred and added to the fits.
Finally, the 100$^{th}$ line after an injection block, dominated by the diffuse
background signal integrated over~4500 TDI line transfers, was included in the fits.
This line represents the lowest signal level in the fits.
 
The calibration datasets were fitted with the \ctiPixel model employing
standard non–linear least-squares minimisation using a Levenberg–Marquardt
Algorithm (LMA) implementation. 
The fits were weighted by the inverse of the sample measurement variance,
derived from a robust determination of the scatter of the input data.
Initially the most problematic aspect of the fits were found to be the large
excess residuals often seen in the first sample of the trails. Investigating this issue, evidence  
of charge blooming from the charge injection
line into the neighbouring pixel was
found in the case of red devices (RP and RVS CCDs), especially at the highest commanded injection voltage levels.
To mitigate this effect, the flux values of the first sample in the case of
the red devices has been corrected using the signal measured in the second
pre-scan sample, i.e.~that nearest the image section at the opposite end of the
serial register. This of course assumes that any charge bleeds at the same rate
forwards and backwards and affects only the pixels adjacent to the image
section. In the case of the blue variant detectors (AF and BP CCDs)
no evidence of any
significant level of charge bleeding was found, and no correction of the
signal in the first trails sample has been attempted.

\section{Fit results}
\label{sec:results}

\subsection{Initial calibrations}
\label{subsec:initialcals}

Visual inspection of the post-scan charge release trails clearly indicates the
need for multi-component decay curve modelling, cf.~deferred charge models as
sum--of--exponentials in \cite{2005ITNS...52.2664H} and references therein,
owing to the presence of trap species of different characteristic charge release
timescales.

\begin{figure}
	\centerline{\includegraphics[scale=0.35]{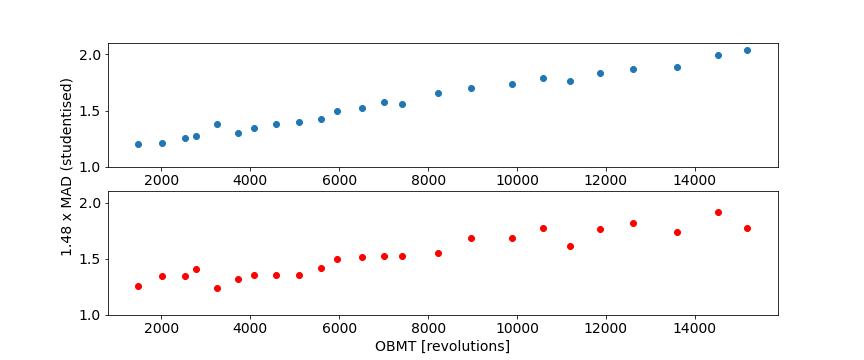}}
    \caption[]{Time evolution in the Gaussian--scaled median absolute deviation
    in unit--weight residual, with time measured as On--Board
    Mission Time  for initial 2--trap species calibrations fitting every device and epoch
    independently at an assumed constant focal plane operating temperature. 
    The top panel is for all blue (AF and BP) devices, the bottom panel for RP devices.}
    \label{fig:qevol-2species}
\end{figure}

Initially we chose a model with two trap species as a baseline to test the
\ctiPixel implementation and associated calibration procedure for all~AF and~XP
devices. Each individual device and epoch was independently calibrated with a
model having~7 free parameters: a single charge volume confinement 
parameter~$\beta$ common between the trap species, plus a trap density $\rho$,
trap emission timescale $\tau_e$ (at an assumed constant nominal operating
temperature in the focal plane) and trap capture cross-section parameter
$\alpha$ for each of the two trap species. The results of these initial
calibrations were rather unsatisfactory, with poor goodness--of--fits and clear
evidence of systematic errors increasing with mission time. 
This is summarised in Figure~\ref{fig:qevol-2species} which shows the robustly estimated unit--weight
residuals of the entire calibrating dataset as a function of mission time,
combined for all blue (AF and BP) and red (RP, but not including RVS) devices. The typical residual is
$\approx1.5$ in both cases with a clear evolution as a function of time.

\subsection{Final calibrations}
\label{subsec:finalcalibs}

The poor quality of the fits of the initial calibrations was addressed 
with a series of refinements to the \ctiPixel model and the fit configuration.
Specifically, two major changes were introduced: we increased the 
number of trap species to three, and implemented a more detailed treatment of
the temperature of the CCDs, initially assumed constant over time and uniform over the field of view. 

The introduction of a third trap species is justified
by the gradual degradation with time in the quality of the initial two~trap species
calibrations, that hints at the emergence of a third defect species of significantly
lower characteristic emission timescale, as radiation damage increased with
mission time. Adding the third trap species to the fit improves the fit results, 
while the increased number of free parameters inevitably leads to
higher levels of correlation and might render their physical interpretation
more uncertain (this can be mitigated as further described below).

\begin{figure}
	\centerline{\includegraphics[scale=0.33]{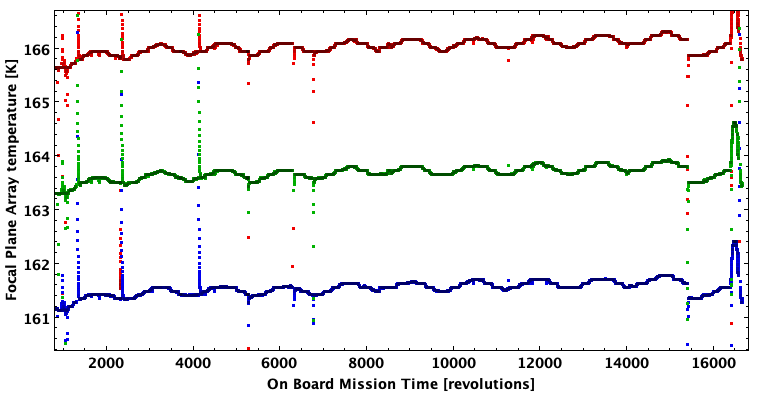}}
    \caption[]{\textit{Gaia} Focal Plane Array temperature variation over the course of
    the ten year mission at three different positions: (blue) row~1~SM; (red)
    row~4~AF5; (green) row~7~RVS. The large variations seen later in the timeline are caused by 
    a series of end-of-life engineering tests.}
    \label{fig:fpatemps}
\end{figure}

In the initial calibrations, the CCD temperature was assumed as fixed at the
nominal focal plane value of $163$~K. This simplification was not
justified on close inspection of the FPA thermal measurements included in the
spacecraft telemetry. A detailed investigation into the thermal properties of
the focal plane revealed variation of many tenths of a degree over time for a
specific device, as well as differences of up to five degrees between CCDs.
Figure~\ref{fig:fpatemps} shows focal plane array temperature telemetry for the
entire mission time line. Three thermistors are positioned on the FPA: with
reference to the FPA device schematic in Figure~1 of~\cite{2016A&A...595A...6C},
these are located on the lower left (row~1~SM), centrally (row~4~AF5) where the
FPA is warmest, and on the upper right (row~7~RVS). The plot shows several
interesting features:
\begin{itemize}
  \item annual variation as solar irradiance changes with the Earth--Sun
  orbital distance;
  \item a secular change from cooler to warmer temperatures from the start of
  the mission to the end;
  \item very short time scale, large amplitude disturbances when operational
  changes cause large fluctuation in power dissipation -- e.g.~the step change near the
  end of mission when Proximity Electronics Module~4 failed on row~3 with the
  FPA subsequently stabilising at an operating temperature lowered by~0.3~K; and 
  \item quantisation at the level of 0.05~K per ADU in the
  on--board analogue--to--digital converters reading the thermistor signals.
  \item large temperature variations induced by the mission end-of-life engineering tests. 
\end{itemize}
The available telemetry was used to create an FPA temperature interpolation map
by device position and mission time with the simplifying assumption of
reflection symmetry to create a 2d temperature model from the limited positional coverage afforded by the FPA
thermistors.

The incorrect assumption of a fixed temperature introduces significant scattering
in the values derived for the fits of the parameter $\tau_e$, the characteristic emission time of a trap species,
due to the exponential in Equation~\ref{equ:tau_emission}.
As mentioned earlier, differences of 5 degrees over the focal plane can introduce 
uncertainties in the estimated release time values of 30\%, for a fast trap species like the ones that 
dominate CTI in the serial register, while a change of 1~degree over the course of the mission
for a specific device can results in estimated trap emission time constants errors of almost 10\%.
In later implementations of \ctiPixel and in the final calibrations the temperature
of each device and its evolution during the epochs has been explicitly defined
in the model. This choice also allows us to fit and directly derive the
fundamental trap species energy level parameter $E_t$, instead of the release
timescale $\tau_e$.

\begin{figure*}
	\centerline{\includegraphics[scale=0.6]{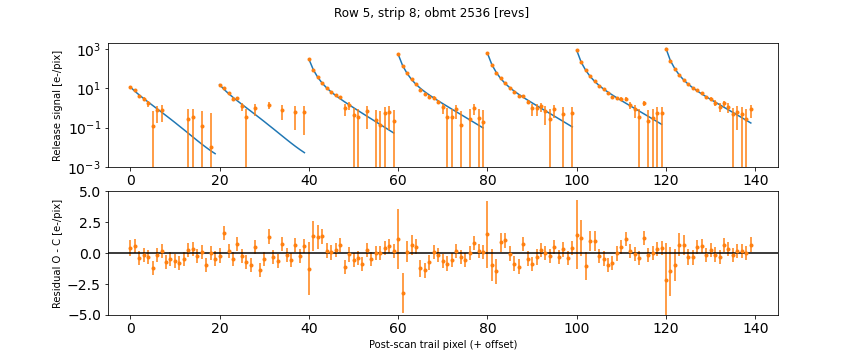}}
	\centerline{\includegraphics[scale=0.6]{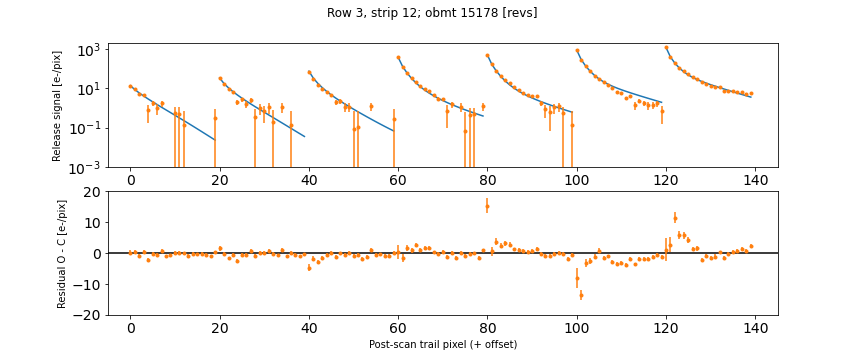}}
    \caption{Calibration data (orange points with measured uncertainties), final
    models (blue lines) and residuals for 3 trap species \ctiPixel models for two device/epochs:
    a typical calibration for row 5, strip 8 (=AF5) near the start of the
    mission at OBMT 2536 revolutions (upper panels); and a poor example for row
    3, strip 12 (=AF9) towards the end of the mission. Calibration data consists
    of seven individual postscan trails, measured over a large range in injected
    signal level, and having a length of 20~pixels each; they are shown
    concatenated in these plots for convenience.}
    \label{fig:cal-3species}
\end{figure*}

\begin{figure}
	\centerline{\includegraphics[scale=0.38]{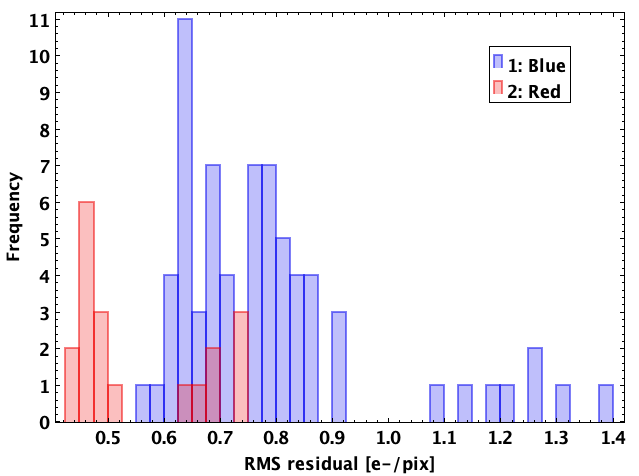}}
	\centerline{\includegraphics[scale=0.44]{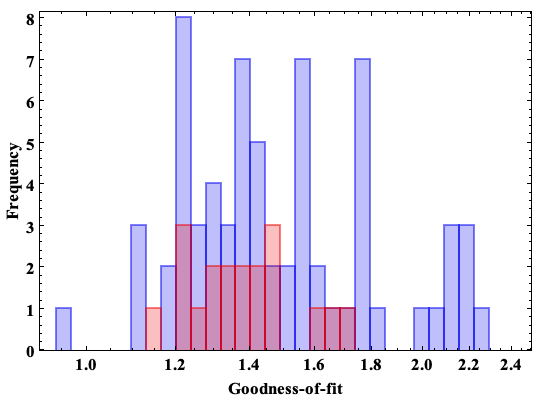}}
    \caption{Residuals (upper panel) and reduced chi--squared goodness--of--fit
    (lower panel) histograms of~87 out of~88 (i.e.~excluding AF9 on row~3) 3
    trap species \ctiPixel final calibrations, subdivided into blue (AF and BP) and red variant (RP and RVS)
    samples (see text).}
    \label{fig:qual-3species}
\end{figure}

\begin{figure}
	\centerline{\includegraphics[scale=0.35]{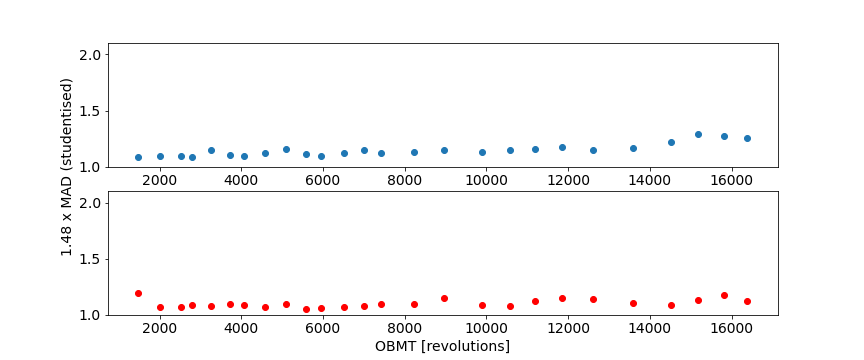}}
    \caption[]{Time evolution in the Gaussian--scaled median absolute deviation
    of the unit--weight residual with time measured as OBMT 
    for the final 3--trap species calibrations (cf.~the same for initial
    2--trap calibrations assuming fixed temperature in
    Figure~\ref{fig:qevol-2species}).}
    \label{fig:qevol-3species}
\end{figure}

Additional attempts to improve the fits by allowing a per--species 
charge volume confinement parameter $\beta$ were
limited in success. After some experimentation in configuration we
settled on a three trap species model including per--species $\beta$ but with
the additional constraint of fitting for a single set of $E_t$, $\alpha$ and $\beta$ 
common across all epochs on a device--by--device basis. The
motivation  was that for a given species we expect those trap parameters to
be constant while the trap density $\rho$ is free to evolve with time.
Figures~\ref{fig:cal-3species} to~\ref{fig:qevol-3species} illustrate the
results for the 3~trap species fits. 
Improvements were evident in the RMS and goodness--of--fit
histograms while the time
evolution of the robustly estimated Gaussian--equivalent unit--weight
calibration residual (in comparison with the same for the 2~trap, individual
epoch calibrations shown in Figure~\ref{fig:qevol-2species}) was flattened
significantly.

\begin{figure}
	\centerline{\includegraphics[scale=0.45]{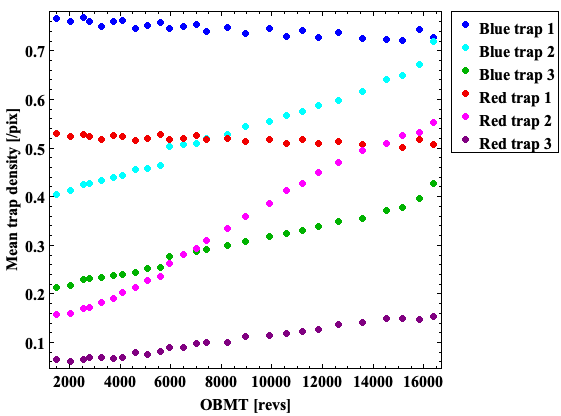}}
    \caption[]{Evolution of the mean trap density with time (measured as On--Board Mission Time (OBMT), 
    in units of the satellite revolution
    period of 6~hours) for the Blue (AF and BP) and Red (RP) CCDs, for the final 3--trap species calibrations.}
    \label{fig:densityevol-3species}
\end{figure}

In Figure~\ref{fig:densityevol-3species} we show the time evolution in mean trap
density for the~3 individual trap species modelled in the blue and red variant
devices. We see clear evidence for strong increase in the density of the higher
energy level traps (i.e.~those with the longer timescales) with time, while the
lowest energy level trap species remains relatively constant in both cases.

\begin{figure*}
	\centerline{\includegraphics[scale=0.6]{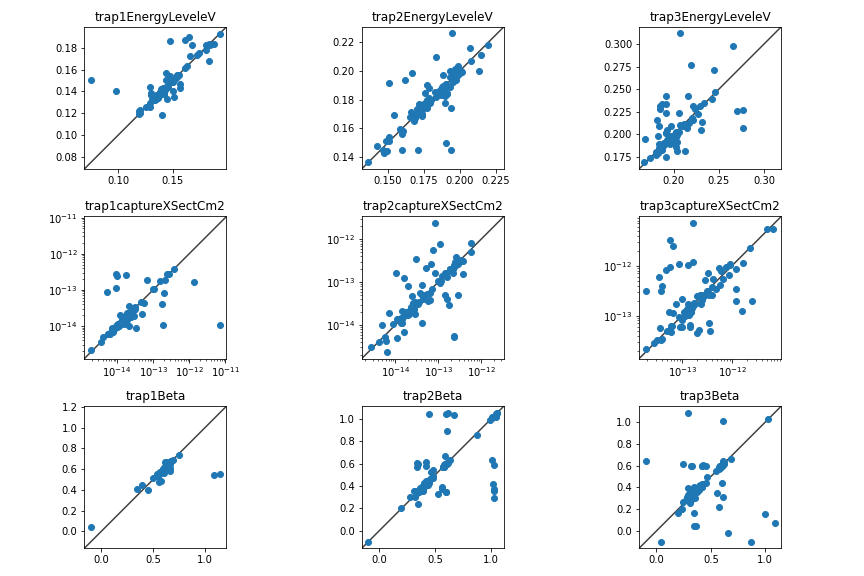}}
    \caption{Comparison of fitted trap parameters between two independent
    calibration runs combining every odd or even epoch's data set
    from the full time--ordered mission set (see text).}
    \label{fig:odd-v-even}
\end{figure*}

In order to investigate the robustness of our results we subdivided the
measurement data into two sets covering the mission timeline, taking
every odd or even epoch in the time--ordered sequence to provide two independent
input data sets of the same size (13~epochs) covering the same overlapping time
range. Figure~\ref{fig:odd-v-even} shows $y=x$ comparison plots for those trap
parameters fixed across the epochs, which we expect to be the same within the uncertainties
propagated through the weighted fits from the measurements, for the three trap
species. The agreement is good for the dominant first species, fair for the
second and somewhat less so for the third, although for many devices the
agreement is reasonable across all trap types. This gives us some
confidence that the trap parameters derived in the final calibrations 
have some useful level of physical reality
despite the likely limitations of the fitting procedure. These include the
arbitrary initialisation of the trap parameters in the non--linear least--squares
optimisation, the possibility of hitting local minima in that process, and the
high level of correlation between some parameters as expected from inspection
of the model as formulated in Section~\ref{sec:gaia_cti_model}.

\begin{figure}
	\centerline{\includegraphics[scale=0.27]{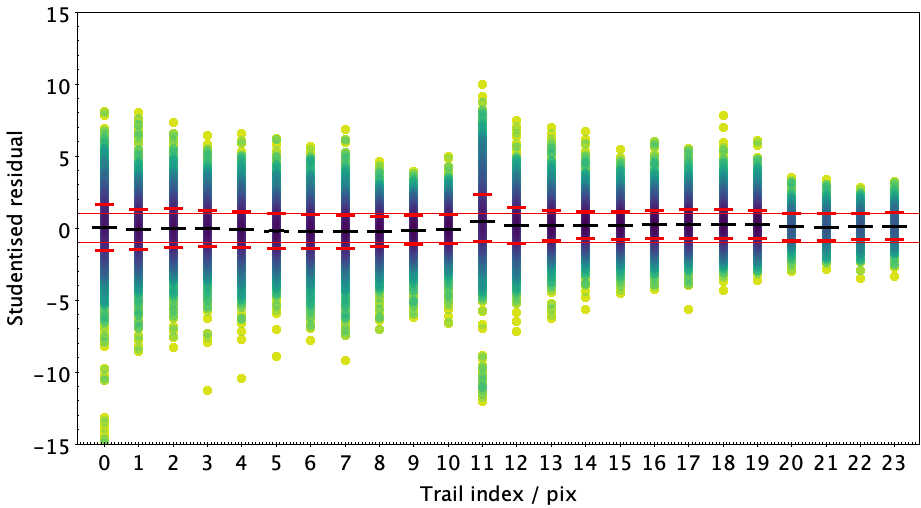}}
	\centerline{\includegraphics[scale=0.27]{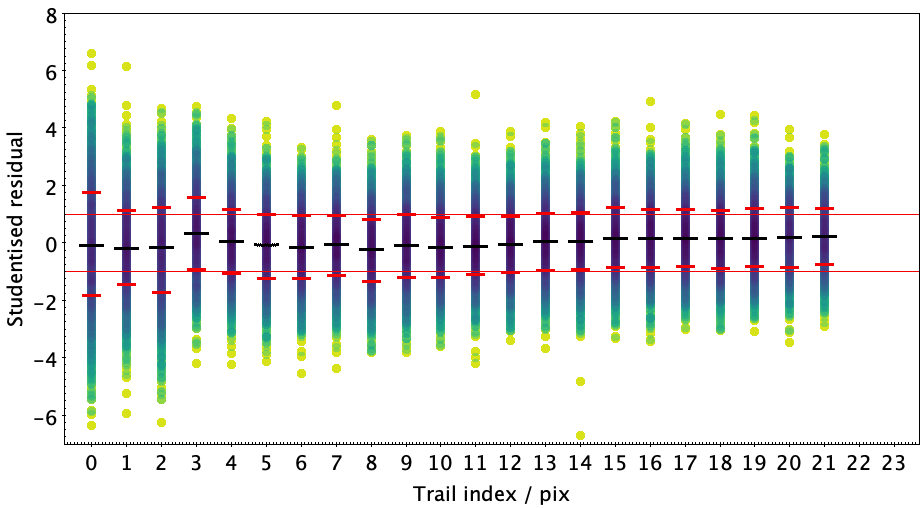}}
    \caption[]{Studentised (unit--weight error) residuals as a function of
    trail position for all 3--trap species calibrations: (upper panel) blue
    variant (AF+BP) devices; (lower panel) red variant (RP) devices. Thick
    dashed lines indicate 16\%, 50\%, 84\% (red, black, red) quantiles
    respectively corresponding to distribution median~$\pm1\sigma$. The thin red
    lines show the $\pm1$ expected position assuming normally--distributed
    uncertainties.}
    \label{fig:trailuwe}
\end{figure}

The goodness--of--fit distribution for the 3~trap species fits
(Figure~\ref{fig:qual-3species}) indicates that the final model is a close
representation of the data, although the median value is $\approx1.5$ rather
than unity as might be expected under the assumption of normally--distributed uncertainties. 
In order to examine potential systematic effects in more detail,
in Figure~\ref{fig:trailuwe} we show scatter plots of the studentised
residuals of the calibrations as a function of sample index in the postscan
trails for
the two CCD variants. These show evidence that the model is a little poorer
in the very first pixels of the release curve following the
injected signal block.
This tendency of the \ctiPixel model to follow less effectively the initial
shape of the decay curves appears across all levels of the injected charge block so
cannot be plausibly explained as poorly mitigated charge diffusion from large
charge packets in the last pixel before the postscan samples. Otherwise, the
only systematic effect clearly present is the residual bias non--uniformity
discontinuity between samples~11 and~12 (blue), and~8 and~9 (red), which results
from the presence of the readout `freeze' during the serial scan
(Section~\ref{sec:trails_fits}).

\section{Discussion of results and insights in the physical nature of the radiation damage}
\label{sec:discussion}

The physically motivated representation of radiation damage
in \ctiPixel is of great advantage in the interpretation 
of the fits results, as it provides a direct link of the model parameters
to the nature of the defects accumulated  
on \textit{Gaia} CCDs over the course of the mission.
At the same time, there are limitations with our analysis that should not be overlooked.
The first one is that, while the results
clearly show that the damage is composed of several defects types, with
distinct charge release times,
only a limited number of post-scan pixels are available to fit the charge trails.
This in turn limits the ability to constrain the fitting parameters, the release
times (or trap energy levels) in particular, and can cause degeneracy between the trap parameters and
possibly over-fitting, in case too many additional species are added to the model.
Even more importantly, the simplifying assumption that 
the damage is composed of a small number of trap species with distinct
energy levels might not be a realistic representation of the
dynamical processes involved in the formation of defects in the \textit{Gaia} space
environment, as discussed below. Finally, as already mentioned earlier, 
the model doesn't appear to fit the very first trail samples perfectly,
an effect that cannot be convincingly explained by 
an incorrect bias subtraction or by injected charge blooming from neighbouring pixels.

The clearest outcome of fitting the serial calibration datasets 
with \texttt{CtiPixel} is the emergence of new trap species 
in addition to the manufacturing Oxygen defect (the A-trap) 
that dominated the serial CTI of the  \textit{Gaia} CCDs at launch.
As can be seen in Figure~\ref{fig:densityevol-3species}, the density of the 
radiation-induced traps shows a general gradual increase, while small step changes
can also be observed shorty after Coronal Mass Ejections directed towards Earth.
Another significant step increase in serial CTI was observed in a dedicated calibration 
dataset collected after an engineering CCD annealing activity, carried out after the
final science observations were acquired. This result is discussed in Appendix~\ref{appendix:eol}.

\begin{figure}
  \centerline{\includegraphics[scale=0.36]{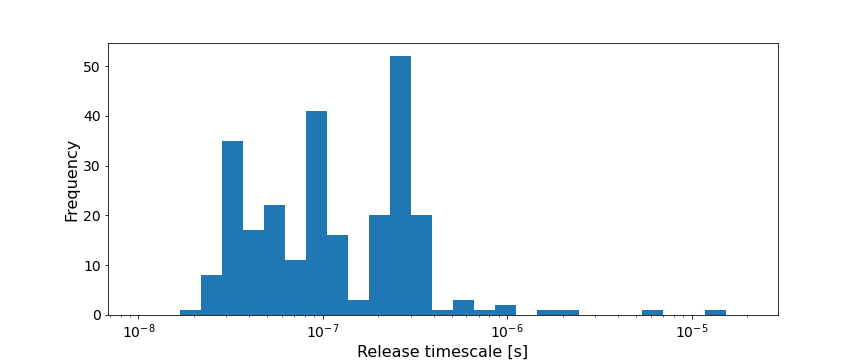}}
   \caption[]{Distribution of trap release timescale values from the final 3--trap species calibrations,
   calculated from the best fit trap energy values of the calibration activity at epoch OBMT = 8224 revolutions.
   Three well defined peaks for the different species of the \texttt{CtiPixel} model 
   are present, including the fastest manufacturing species that dominated the CCD defects at launch
   and two slower radiation-induced traps. 
   A tail of slower emission times can be observed for the third species. These can be interpreted
   as emerging defects still in an unstable configuration, similarly to the radiation-induced 
   defects measured on board the Euclid mission CCDs.}
\label{fig:trap_release_hists}
\end{figure}

The new species generated by space radiation present higher energy levels and 
therefore longer emission timescales compared to the A-trap. Their peaks in 
the distribution of the traps release times can be
clearly identified in Figure~\ref{fig:trap_release_hists}.
The intermediate trap species is characterised by an
energy level of approximately 0.20 eV below the conduction band 
and emission time constant of $1.25 \times 10^{-7}$~seconds at the CCDs operating temperatures, 
suggestive of a possible association with the multi-vacancy defect.
The slowest species peaks at approximately $6 \times 10^{-7}$~seconds;
interestingly, its distribution appears to have a tail of longer release time values, 
extending as long as $\sim 10^{-5}$~seconds.

The origin of the wider distribution of characteristic energies and emission values of the third species 
could be in part attributed to 
poorly constrained best fit parameters due to the relatively low density of this defect, 
especially early on in the mission. An alternative, plausible explanation is 
that the observed continuum is instead intrinsic to this type of defect at the time of the measurements,  
when the newly generated traps had not yet formed a stable bond with
impurities or other vacancies in the lattice: 
in this case, the wide range of derived energy level values reflects their transitional state and the 
dynamics of the damage accumulation. A similar scenario has been experienced
by the VIS CCDs \citep{2024arXiv240513492E} on board ESA's Euclid mission
\citep{2024arXiv240513491E}.
The analysis of the damage experienced by the VIS CCDs during Euclid's first year in space,
measured using the trap pumping calibration technique --- also known as pocket pumping
\citep{2001sccd.book.....J,2014ITNS...61.1826H} --- showed a  
continuum of emission values for the newly developed traps
\citep{arXiv:2407.01268}, just as seen in the \ctiPixel model fits of \textit{Gaia}'s damaged CCDs.

\begin{figure}
  \centerline{\includegraphics[scale=0.49]{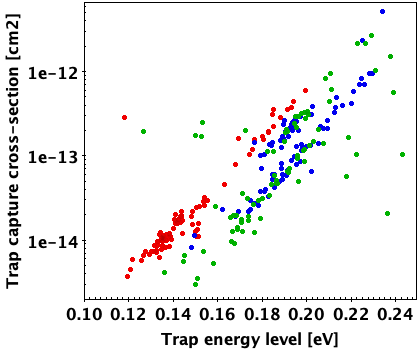}}
   \caption[]{Scatter plot of trap capture cross--sections [cm$^2$] versus
   energy level [eV] for the 3 trap--species model fits to the entire \textit{Gaia}
   mission data set. The red, green and blue dots are relative to the first, second and 
   third trap species in the \ctiPixel model fit.}
\label{fig:sigmasVets}
\end{figure}

The clustering of the trap parameters for the distinct species can also be seen in the 
scatter plot of Figure~\ref{fig:sigmasVets} for the final 3--species fits. 
In this figure the fitted trap parameters~$\alpha$ have
been translated into a corresponding capture cross--section for each device at
the appropriate operating temperature (see Equation~\ref{equ:xsection}).  
While the model parameters are evidently correlated, the plot
illustrates the presence of at least two clearly distinct trap species populations, clustering 
at specific energy levels and cross section values.

The evolution of the measured mean damage of
Figure~\ref{fig:densityevol-3species}, seems to suggest a slight decline in the
number density of the A-type manufacturing species, that dominated the defects at launch. 
This apparently surprising result is probably
simply due to the degeneracy between the species' parameters 
when a third defect type is added to the fits.
It is indeed unlikely that the observed trend could alternatively be caused, 
at least in part, to the physical process of annealing. 
Thermal annealing, resulting in the rearrangement of defects within the silicon 
lattice, happens quite quickly at high CCD temperatures, but in the cold environment of the  \textit{Gaia} focal plane 
any annealing process would be very slow, and therefore only contribute to the rearrangement 
of a very small fraction of defects.

\begin{figure*}
	\centerline{\includegraphics[scale=0.45]{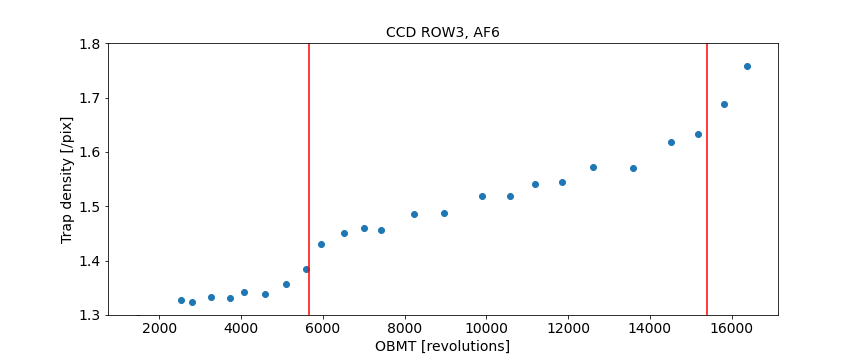}}
    \caption{Evolution of the total density of traps for the CCD in row 3 and AF strip 6.
    The observed trend is caused by a gradual increase in damage caused by cosmic rays and sudden step jumps in 
    correspondence to solar flares event, shown as the red lines for the class X8.2 flare of September 2017 and 
    as the start of the intense period of solar activity of Spring-Summer 2024.}
    \label{fig:densityRow3AF6}
\end{figure*}

Averaging the trap densities over all devices as shown in Figure~\ref{fig:densityevol-3species} risks hiding 
the finer details of the accumulation of defects in the \textit{Gaia} devices. 
To more clearly reveal all aspects of the damage evolution, 
we present the example of CCD \textit{(ROW3, AF6)}, positioned in the third row
and sixth strip of AF devices in the focal plane in Figure~\ref{fig:densityRow3AF6}.
The Figure shows the combined number of defects per pixel
of the 3 trap species as a function of the number of revolutions since launch, as derived
from the \ctiPixel best fits.
The general trend is dominated by a gradual accumulation of defects caused by the impact
of Galactic cosmic rays on the detectors. 
As in the parallel case \citep{2016SPIE.9915E..0KC} the serial damage
also shows step increases in data taken immediately following the strongest 
solar flares and coronal mass ejections of the last solar cycle. 
An interesting difference in the measured damage evolution between the two readout directions 
is observed in the magnitude of the step changes after
the major flaring events in September 2017 and in the spring of 2024. 
The 2017 X8.2 class flare had the largest
effect in the measured CTI in the image section (parallel readout direction), 
while in the serial register it resulted in a modest step increase of the total trap density.
The opposite appears to be true after the 2024 period of strong solar activity.
On further investigation, this surprising result has been attributed to the 
sudden temperature variation in the \textit{Gaia} focal plane recorded in May 2024,
that contributed to the changes in measured CTI. The colder operating temperatures have reduced 
its impact in the parallel direction, 
while in the serial register the readout losses have increased significantly. 
At this cooler setting the traps emission timescale
have shifted to longer values (see Equation~\ref{equ:tau_emission}), 
impacting in opposite ways the ability of traps to capture electrons 
in the image section and the readout register, due to their different readout transfer frequencies.

\begin{figure*}
 \centerline{\includegraphics[scale=0.48]{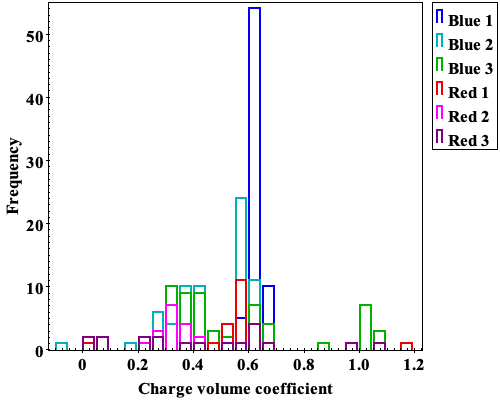}
            \includegraphics[scale=0.48]{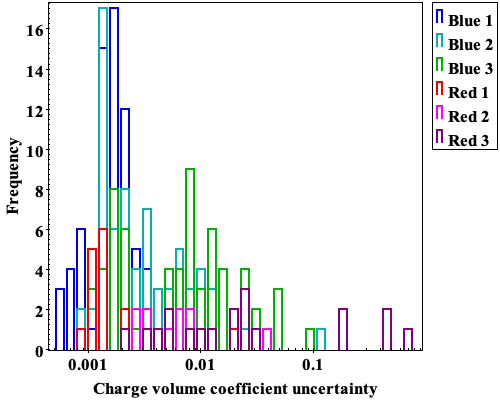}}
 \centerline{\includegraphics[scale=0.48]{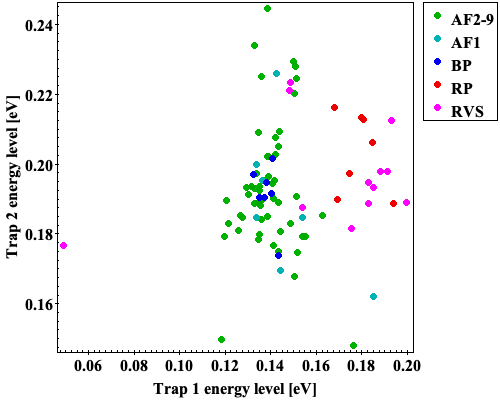}
            \includegraphics[scale=0.48]{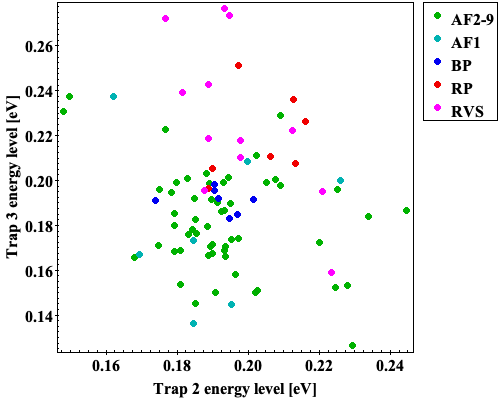}}            
 \centerline{\includegraphics[scale=0.48]{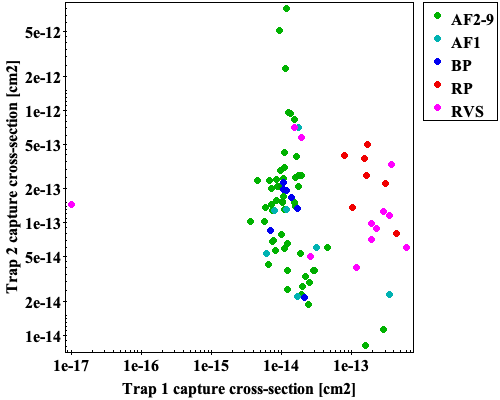}
            \includegraphics[scale=0.48]{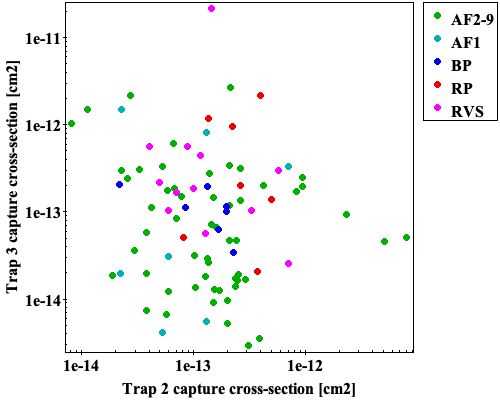}}
 \caption[]{Trap parameters for the final 3--trap species calibrations plotted
 as histograms (charge volume confinement exponent~$\beta$; top row) and
 scattergrams (trap species energy level~$E_t$, middle row; trap capture
 cross-section parameter~$\alpha$, bottom row) for the full calibration
 datasets.}
 \label{fig:finalcals}
\end{figure*}

Figure~\ref{fig:finalcals} shows fits parameters resulting from the fits of the entire set of
serial calibration data available during the mission in the form of scatter plots and histograms.
Outlying points from fits with Goodness of Fit (GoF) on the tail of the
distribution are not shown in the Figure. In particular, the \textit{(ROW3, AF9)} CCD is not well
modelled by \texttt{CtiPixel}. This device presents a defect in column \textit{AC}=1977, that could be 
responsible for the excess charge in the first post-scan pixel due to deferred charge that is causing 
poor fits of the charge trails, especially at high charge injection levels. We notice that \textit{(ROW3, AF9)} 
is a 'twin' CCD, one of 10 pairs of devices present in the \textit{Gaia} focal plane (we use the term 'twins' to indicate
a pair of CCDs manufactured from the same 
silicon wafer; the \textit{(ROW3, AF9)} pair is \textit{(ROW2, AF4)}, whose fit presents a typical GoF value). 
Appendix D in \cite{nickPSFInPrep} presents a list of \textit{Gaia} CCD twins and information on the device types.

With respect to $\beta$, the charge volume 
coefficient, the parameter that in \ctiPixel links the number of signal electrons with the confinement volume 
(the physical volume filled by the charge packet in a pixel), 
the measured distribution for the third trap species appears to separate from the other types of defects.
This result could indicate a different 
physical distribution of this trap species within the pixels, and could indeed suggest that
these defects are still migrating in the lattice and have not yet bonded into a stable defect.

Another interesting result to note in these plots is the different parameter space occupied
by the red variant (RP) devices with respect to the blue CCDs, in terms of both
the energy levels (and hence release timescale) and the traps cross sections.
This result is not well understood. As shown in
Figure~\ref{fig:densityevol-3species}, the number of manufacturing defects in the serial register was lower in red
devices at launch, while the rate of traps increase appears similar for all devices (see also \cite{10.1117/12.2562162}).
This hints to possible differences in the distribution of the impurities during the manufacturing process of 
thicker and thinner devices, even if both the pixels architecture and the serial readout process
are common to all devices.

The fits to the calibration data sets occasionally show systematic effects in
the residuals in the very first pixels of the charge trails, as mentioned
previously.
Effects such as an inaccurate subtraction of the CCD bias, or 
charge blooming from the high injected signal flux in the last pixel of the serial register 
seem unlikely causes of the residuals. 
For instance, such a blooming effect is not detected in the first pre-scan pixel.
The most plausible explanation of the poor fits 
is an inadequacy in the \ctiPixel implementation, possibly too simplistic, 
in modelling the full range of the trap emission timescales. 
Once again, the intrinsic limitation of a
charge distortion modelling that allows only a small number of trap species
with distinct trap parameters could be the cause of the limited success of
\ctiPixel to fully account for all the observed CTI effects. 

The \ctiPixel model was developed with the main goal of characterising the defects responsible for the 
effects of serial CTI in \textit{Gaia} CCDs and to measure the evolution of radiation damage at L2.
Trying to apply the model, calibrated using engineering data, to science observations is a challenging endeavour 
for a number of reasons. For instance, the PSF needs to be sampled fully in 2D for all predicted sources. 
Processing of science data would require predicting the scene of each science window 
to account for any other signal that might affect the history of the traps and their occupancy level, 
including background variations. The scene outside the PSF core would inevitably be poorly 
characterized, as the mid-to-far PSF wings are not well-covered by calibration data. This would be problematic 
in the serial CTI context, where the wings in the Across Scan direction are most important.
A proper application of the serial CTI model would also need to account for other CCD response effects, such as the
column response non uniformity, when realising the predicted scene. The readout sequencing of science data would be 
critical in the application of the CTI model, but this aspect
might not have been well-constrained by the calibrations using only engineering sequences.

In the attempt to apply the calibrated \ctiPixel model
for science data processing, the PSF model was used in combination with the background modelling and 
other auxiliary calibrations such as the bias non-uniformity corrections to estimate the science inputs, 
applying the \ctiPixel algorithm to derive a complete model of a selected sample of observations, including serial CTI.
The results of the application were only partially satisfactory, as 
the calibrated model reproduced qualitatively 
the serial CTI signature observed in the science windows but could not fully model its effects under 
all conditions. In particular, the model struggled to reproduce the observed CTI effects for very bright sources and
in the case of high local background, where the diffuse distribution of charge over a device mitigates the losses 
due to traps.

The result points to possible limitations with the accuracy of the \ctiPixel model and its calibration 
with the sole use of the dedicated serial CTI charge injection activities. It is possible,
for example, that the limited number of post-scan pixels for which the charge trail flux values are recorded 
 is insufficient to identify trap species of longer time emission constants, that could be present
 in the serial register and affect the readout.
A second intrinsic limitation of the calibration using post-scan trails is that only the integrated effect of the damage
is measured, based on which the \textit{average} density of traps in the serial register can be derived; in reality, an accurate model 
for the treatment of serial CTI for stars and other point sources requires the estimate of the 
density of traps in each pixel, to account for its spatial non-uniformity (both of the initial manufacturing defects and 
of the traps accumulating due to radiation damage).
The assessment of the accuracy of the \ctiPixel model,
when applied to images affected by radiation damage,
could have been hindered by other systematic effects that were not fully calibrated at the time science validation
was attempted. This is for example the case of the brighter/fatter effect, that can result in image residuals of
the same order as the ones caused by serial CTI.
Furthermore, it is worth noting how the PSF model 
does not actually represent the purely linear part of the signal, as real science observations, that
present a certain level of serial CTI depending on their signal level and the specific AC position on the detector, 
are used for its calibration. The attempted science tests were therefore not ideally suited to 
reproduce the exact scenario for which \ctiPixel was designed and developed.

A critical aspect already mentioned earlier that was experienced 
when attempting the processing of the test science data sets with \ctiPixel
is the high demand of computation resources needed when modelling the image readout at the pixel
level. After these initial, and not fully successful test cases, it was concluded that the integration
of the \ctiPixel task into the  \textit{Gaia}'s Image Parameter Determination software (IPD)
for the treatment of the CTI distortion effects would not be feasible, as it would exceed the 
accessible machine processing time, 
considering the sheer volume of \textit{Gaia} observations and the cyclic nature of the data reduction.
It is also worth noting that, in the \textit{Gaia} context, the AC location bias caused by serial CTI are not as critical 
as the CTI bias in the parallel direction, since the AC source locations are not used in the AGIS astrometric solution \citep{2016arXiv160904303L},
and are only indirectly employed in the attitude and calibration solutions. 
The efforts with the characterisation and the analysis of CTI were therefore ultimately redirected to a project to report passively on the observed biases, 
without attempting any processing of science data.

\section{Conclusions}
\label{sec:conclusions}

The 106 CCDs of ESA's \textit{Gaia} spacecraft have been exposed to the particle irradiation of the space environment 
during the course of the 11-year mission.
Dedicated serial CTI calibration engineering activities, performed periodically on board the spacecraft, 
have been used to measure the  effect of radiation damage on the detectors, and to monitor its evolution.
A pixel-based, physically motivated \ctiPixel model that adopted the Shockley-Read-Hall theory of charge traps 
was developed to analyse the serial CTI calibration data sets. 
The results of the investigation have shown that alongside the manufacturing 
defects already present at launch, new trap species caused by particle irradiation have emerged. 
The newly generated defects are characterised by longer charge release
timescales compared to the original manufacturing defect. The intermediate trap has a measured 
energy level consistent with the known value of the multi-vacancy 
defect, while the third species presents a tail in its release time distribution,
suggestive of a dynamical, transitional state of some radiation-induced traps, 
with defects not fully settled into their final configuration.
The evolution of the radiation-induced serial damage appears linear, 
with the exception of calibration epochs following strong CME 
directed towards Earth like the ones occurring in 2017 and 2024, 
that present step-increase in damage level. 
The latest epochs also show the effects of temperature instabilities 
of the focal plane on the measured serial CTI, as well as 
a further step increase caused by the engineering CCD annealing test carried out
after the end of \textit{Gaia} science operations.

\begin{acknowledgements}

This work has made use of data from the European Space Agency (ESA) mission
{\it Gaia} (\url{https://www.cosmos.esa.int/gaia}), processed by the {\it Gaia}
Data Processing and Analysis Consortium (DPAC,
\url{https://www.cosmos.esa.int/web/gaia/dpac/consortium}). Funding for the DPAC
has been provided by national institutions, in particular the institutions
participating in the {\it Gaia} Multilateral Agreement. The \textit{Gaia}\
mission website is:
\url{http://www.cosmos.esa.int/gaia}.

This work was financially
supported by the European Space Agency (ESA) in the framework
of the Gaia project; and the United Kingdom Space Agency (UKSA)
through the following grants to
the University of Edinburgh and the University of Leicester, ST/S000976/1, ST/S001123/1.
The authors gratefully acknowledge the use
of computer resources from MareNostrum, and the technical expertise and assistance
provided by the Red Española de Supercomputación at the Barcelona
Supercomputing Center, Centro Nacional de Supercomputación. 
Diagrams were produced using the astronomy-oriented data handling and visualisation
software TOPCAT \citep{taylor}.

\end{acknowledgements}

\bibliographystyle{aa}
\bibliography{refs}

\nolinenumbers
\begin{appendix}
\section{End of Life test}
\label{appendix:eol}
\nolinenumbers

Gaia scientific observations ended on 15 January 2025. The spacecraft was moved 
from its position around L2 into a 
retirement heliocentric orbit with a final burn of its thrusters on 27 March 2025, and passivated.
Before this final procedure, a series of engineering and technology tests were conducted to investigate 
the instruments and spacecraft
behaviour, to improve calibrations and inform future missions. Among these tests, of importance for 
radiation damage and CTI studies was the CCD radiation damage annealing test.

During the annealing procedure, on the 6th of March 2025
the temperature of the focal plane was increased to room temperatures (in a range
between 27 and 38 degrees Celsius across the focal plane) for the duration 
of four spacecraft revolutions (corresponding to approximately a full day).
The expected outcome of the test was a partial reduction of the measured CTI, in particular 
in the parallel direction, as most traps in the image area 
have been induced by on-orbit irradiation at cold temperatures.
In the serial register, where most of the traps even at the end of the mission are still the original 
manufacturing defects already present at launch a lower improvement in CTI was expected, 
as the O-V (oxygen-vacancy) defect is known to be stable below 150~C \citep{2001sccd.book.....J}. 

Two days later, at OBMT revolution 16638, a final serial CTI calibration activity was conducted,
allowing the investigation of the effect of the bakeout on the Gaia devices.
This End-of-Life (EoL) serial calibration activity was conducted following the same procedure as all
the previous pre-annealing epochs, injecting charge in blocks at a number of different injection levels.
The dataset collected from the final epoch could therefore be
processed following the same procedure already 
described for the earlier datasets,  but limiting the fit to the data of the single EoL epoch.
With this choice, any change in the traps parameters could be measured and compared
to previous epochs, specifically in regard to the density of the species and any 
rearrangement of their energy levels.
Two major changes were measured following the bakeout: in the parallel direction, 
a reduction of the effects of CTI of approximately 30\% was observed, 
as derived from the analysis by Gaia payload experts (private communication); inversely, and maybe unexpectedly,
in the serial direction a reverse-annealing effect was seen, with an increase in the
measured density of defects. 

\begin{figure}
  \centerline{\includegraphics[scale=0.52]{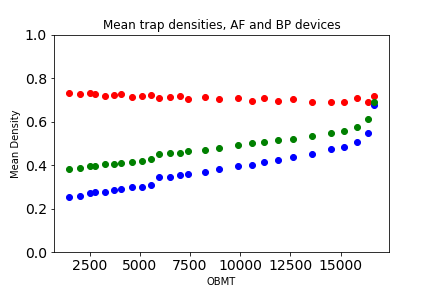}}
    \centerline{\includegraphics[scale=0.52]{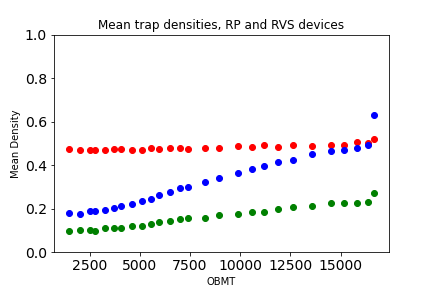}}
   \caption[]{Time evolution of the mean densities of the three trap species over the 
   entire mission, for all blue devices (AF and BP, top panel), and for the red devices 
   (RP and RVS, bottom panel). A clear step in the density of the radiation-induced traps is
   seen for both the blue and the red devices, while the density of the manufacturing trap appear unaffected 
   by the annealing procedure.}
\label{fig:meanDensities}
\end{figure}

The clear evidence of an increase in the trap densities after the CCD annealing test is observed 
from the evolution of the best fit parameters, as shown in Figure~\ref{fig:meanDensities}, for the blue
and red devices (top and bottom plots). In detail, while the density of the manufacturing defects (red circles) 
appears to be stable after the bakeout, a clear step is seen for the two radiation-induced species, 
for both types of devices.
The outcome of the bakeout procedure seems to point to a partial
rearrangement of the defects caused by space radiation, 
with the formation of new stable defects with a characteristic release timescale that is
able to affect the readout at the serial clocking frequency.
On the other hand, for the manufacturing defects, already present at launch and that formed at room 
temperatures, the annealing test temperature wasn't sufficiently high 
to affect their bonding with the Oxygen impurities (bakeout temperatures of approximately 600~K are indeed needed to anneal the O-V defect).
This result is reminiscent of the outcome of the annealing procedure that was 
attempted on board NASA's Chandra X-ray observatory,
following the sharp increase of CTI measured in the 
Front Illuminated advanced CCD Imaging Spectrometer (ACIS, \cite{2003SPIE.4851...28G}) 
shortly after launch, caused by the exposure to soft proton. 
The Chandra focal plane was heated from the nominal operating temperature of -100~C to +30~C, in an attempt
to alleviate the observed CTI increase, but the process resulted in a further increase of CTI, 
attributed to the formation of new defects caused 
by Carbon impurities in the ACIS CCDs (\cite{article}, \cite{MONMEYRAN201623}).
   
\begin{figure*}
  \centerline{\includegraphics[scale=0.55]{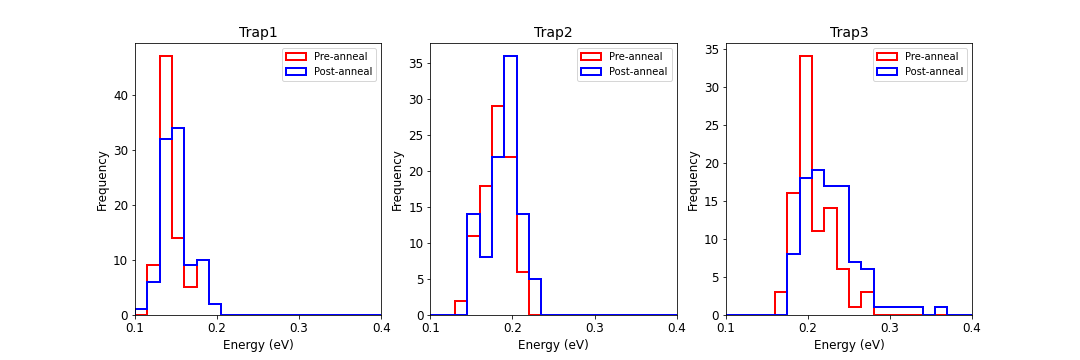}}
   \caption[]{Distribution of the characteristic energy levels of the three trap species, from the best fit
   from the pre and post-annealing epochs (from left to right: first, second and third species).}
\label{fig:annealTrapEnergies}
\end{figure*}

Figure~\ref{fig:annealTrapEnergies} compares the distribution of the energy levels of the three trap species 
derived from the fits of the pre- and post-annealing epochs. While the distributions of the fitted
energy levels of the manufacturing trap (the first species) and the 
intermediate trap species appear similar before and after bakeout,  
the energies of the slower species appear slightly shifted in the EoL epoch,
and with a tail towards higher energies. This finding could be interpreted as an indication of a partial 
rearrangement of these defects to a new configuration after the interval at higher temperatures.

The lines immediately following the injection blocks of the serial calibration activities can inform
on the level of CTI damage in the parallel direction, as their flux, after the appropriate bias and 
background have been subtracted, consist of deferred charge by traps in the image section of the CCDs.
This analysis revealed a reduction of approximately 25-30\% in the amount of deferred charge after 
the EoL annealing procedure.
This outcome is consistent with expectations, as most traps causing parallel readout charge losses have
formed in orbit at cold temperatures, 
and the focal plane bakeout has resulted in the partial annealing of the unstable defects.

\end{appendix}

\end{document}